\documentclass[runningheads]{llncs}

\newif\ifpublish
\publishtrue

\usepackage{amsmath}
\usepackage[T1]{fontenc}
\usepackage[hidelinks]{hyperref}
\usepackage{graphicx}
\usepackage{xspace}
\usepackage{cleveref}
\usepackage{xcolor}
\usepackage{enumitem}
\setlist[itemize]{leftmargin=*}
\usepackage{algorithm}
\usepackage[noend]{algpseudocode}
\usepackage{multicol}
\usepackage{breakcites}
\usepackage[font=footnotesize]{caption}

\newcommand{\codelink}{
    \ifpublish
        \url{https://github.com/asonnino/mysticeti/tree/obelia} (commit \texttt{fe74642})
    \else
        Code available but link omitted for blind review.
    \fi
}

\newcommand{\sysname}{\textsf{Obelia}\xspace}
\renewcommand{\dag}{\ensuremath{D}\xspace}

\newcommand{\para}[1]{\smallskip\noindent\textbf{#1}.}
\newcommand{\change}[1]{\begingroup\color{orange}#1\endgroup}
\newcommand{\require}[1]{\textbf{require} #1}
\newcommand{\algfontsize}{\scriptsize}

\begin{document}

\title{Short Paper: \sysname: Scaling DAG-Based Blockchains to Hundreds of Validators}
\titlerunning{\sysname}

\ifpublish
    \author{
        George Danezis\inst{1,2} \and
        Lefteris Kokoris-Kogias\inst{1} \and
        Alberto Sonnino\inst{1,2} \and \\
        Mingwei Tian\inst{1}
    }
    \authorrunning{G. Danezis et al.}
    \institute{
        Mysten Labs \and
        University College London (UCL)
    }
\else
    \author{}
    \institute{}
\fi

\maketitle

\begin{abstract}
    \sysname improves upon structured DAG-based consensus protocols used in proof-of-stake systems, allowing them to effectively scale to accommodate hundreds of validators. \sysname implements a two-tier validator system. A core group of high-stake validators that propose blocks as in current protocols and a larger group of lower-stake auxiliary validators that occasionally author blocks. \sysname incentivizes auxiliary validators to assist recovering core validators and integrates seamlessly with existing protocols. We show that \sysname does not introduce visible overhead compared to the original protocol, even when scaling to hundreds of validators, or when a large number of auxiliary validators are unreliable.
    \ifpublish
    \fi
\end{abstract}

\section{Introduction}

Blockchains using BFT quorum systems~\cite{sui,aptos,solana} divide time into 24-hour epochs, during which a committee of about 100 \emph{validators}, elected through a Sybil-resistant mechanism~\cite{douceur2002sybil}, often a variant of proof-of-stake~\cite{kucci2021proof}, operates the system using a BFT consensus protocol~\cite{pbft,jolteon,mysticeti}. Their voting power correlates with their stake, allowing agreement on blocks of client transactions. Recent advancements in BFT protocols utilize directed acyclic graphs (DAG)~\cite{narwhal,bullshark,shoal++,mysticeti,sailfish,bbca-chain,gradeddag,cordial-miners,fin,fino}, achieving high throughput (> 100k tx/s) and robustness against faults and network asynchrony~\cite{consensus-dos,narwhal}.

However, these consensus protocols limit operation to approximately 100 validators, sidelining many potential participants---often in the hundreds~\cite{sui-scan}. This exclusion is a sharp contrast to more traditional blockchains like Bitcoin~\cite{bitcoin} and Ethereum~\cite{ethereum}, which engage all participants, and is responsible of key weaknesses of quorum-based blockchains.
First, only the subset of the total stake hold by these validators can be used to decentralize the system and benefit the blockchain ecosystem. Lower-stake players cannot participate in block proposals and typically resort to running full nodes without incentives~\cite{krol2024disc}.
Second the high throughput of DAG-based systems complicates state catch-up for new or crash-recovering validators, who either strain the committee's resources or depend on external unincentivized entities for recovery.

This paper introduces \sysname, an enhancement to DAG-based consensus that increases participation by enabling all stakeholders to sporadically author blocks. It incentivizes these participants to assist recovering validators and integrates seamlessly with existing protocols. However, developing \sysname involves overcoming significant challenges.
(1) \sysname must avoid all-to-all communications between stakeholders as their large number makes it impractical.
(2) It cannot rely on a classic BFT assumption for entities that have less stake and thus less incentive to be reliable. This challenge results from the inherent poor reliability of these slow-stake entities that can be offline for long periods of time. \sysname must ensure that all data they contribute to the chain remains available. Where traditional systems rely on monetary penalties to disincentivise unreliability~\cite{he2023don} by assuming network synchrony, \sysname cannot follow this guidance as it aims to operate in the weaker asynchronous or partially synchronous network model of existing quorum-based protocols.
(3) The final challenge consists in allowing these low-stake entities to participate in the consensus without slowing it down, as this would compromise the major benefit of quorum-based systems.

\sysname addresses these challenges by introducing a two-tier validator system. A core group of high-stake validators proposes blocks as in existing protocols, while a larger group of lower-stake auxiliary validators occasionally authors blocks. Auxiliary validators operate outside the critical path, proposing blocks at a slower pace and only after obtaining a strong proof of availability for their pre-disseminated block. Our implementation and evaluation of \sysname demonstrate that it does not introduce noticeable overhead compared to the original protocol, even when scaled to hundreds of potentially unreliable auxiliary validators.

\para{Contributions}
This paper makes the following contributions:
\begin{itemize}
    \item We present \sysname, a novel mechanism enhancing DAG-based protocols enabling all stakeholders to engage in consensus and incentivizing support for recovering validators.
    \item We demonstrate \sysname's safety and liveness within the same network model as its underlying quorum-based protocol.
    \item We implement and evaluate \sysname on a realistic geo-distributed testbed, showing it adds negligible overhead despite a large number of potentially unreliable low-stake validators.
\end{itemize}

\section{System Overview} \label{sec:overview}

We present the settings in which \sysname operates.

\subsection{Validators selection} \label{sec:validators}
\sysname introduces the distinction between \emph{core validators} and \emph{auxiliary validators}. Core validators are the validators that \emph{continuously} operate the consensus protocol and process transactions. In contrast, auxiliary validators participate sporadically while maintaining a copy of the DAG generated by core validators.
Both core and auxiliary validators are selected using a sybil-resistant mechanism~\cite{douceur2002sybil}, typically based on proof-of-stake~\cite{kucci2021proof}. Core validators are chosen similarly to existing quorum-based blockchains, consisting of roughly the 100 entities with the highest stake or those meeting specific criteria, such as owning a minimum percentage of the total stake~\cite{sui}.  Auxiliary validators include all other stakeholding entities not in the core group, typically numbering in the several hundreds~\cite{sui-scan}, significantly surpassing the number of core validators.  In practice, we expect current full nodes to operate as auxiliary validators.

\subsection{System and threat model} \label{sec:model}
\sysname assumes a computationally bounded adversary, ensuring the security of cryptographic properties such as hash functions and digital signatures. It operates as a message-passing system where core and auxiliary validators collectively hold  $n = n_c + n_a$ units of stake~\cite{saad2020comparative}, with $n_c$ held by core validators and $n_a$ held by auxiliary validators. Each unit of stake represents one ``identity''~\cite{douceur2002sybil}, while each unit held by a core validator signifies one ``unit of voting power'' in the consensus system~\cite{sui-code,sui}. This model aligns with deployed quorum-based blockchains, where core validators possess the majority of total stake ($n_c \gg n_a$)~\cite{sui,aptos,solana}. \sysname makes the following assumptions for core and auxiliary validators:

\para{Core validators}
\sysname works with existing DAG-based consensus protocols, inheriting their assumptions. Specifically, it requires that $n_c \geq 3f + 1$, where $f$ is the maximum number of \emph{Byzantine} core validators that may deviate from the protocol. The remaining stake is held by \emph{honest} core validators who adhere to the protocol. There are no additional assumptions about the network model, core validators operate in the same setting as the underlying DAG-based consensus protocol. Note that most deployed DAG-based consensus protocols are partially synchronous~\cite{dwork1988consensus}, while some blockchains consider asynchronous protocols~\cite{sui-code}. Under these assumptions, \Cref{sec:security} demonstrates that a DAG-based protocol enhanced with \sysname is \emph{safe}, meaning no two correct validators can commit conflicting transactions.

\para{Auxiliary validators}
For auxiliary validators, \sysname adopts a relaxed model due to their lower stake and reduced incentives for resource dedication and reliability. It assumes that at least $t_a \leq n_a$ units of stake are consistently held by honest and active auxiliary validators, regardless of the total number of auxiliary validators. The parameter $t_a$ can be adjusted to balance system liveness (see \Cref{sec:security}) against the minimum participation of auxiliary validators. Auxiliary validators do not communicate with one another and only occasionally communicate with core validators over an asynchronous network. \Cref{sec:security} shows that, under these assumptions, a DAG-based protocol enhanced with \sysname is \emph{live}, ensuring that honest validators eventually commit transactions. Importantly, if the assumptions concerning auxiliary validators fail, safety remains guaranteed.

\subsection{Design goals and challenges} \label{sec:goals}

Beyond ensuring safety and liveness within the same network model as the underlying consensus protocol, \sysname achieves several design goals (discussed in \Cref{sec:design}):
\textbf{Increased participation (G1):} It allows all entities holding stake to author blocks in the consensus protocol, rather than limiting participation to the top 100 validators.
\textbf{Incentivized synchronizer helpers (G2):} \sysname leverages auxiliary validators to assist slow or recovering core validators in catching up to the latest state. This approach incentivizes auxiliary validators to function as full nodes, storing and providing the DAG state to core validators to facilitate synchronization.
\textbf{Generic design (G3):} The design of \sysname is directly applicable to a wide range of structured DAG-based consensus protocols.

\sysname also has performance goals that we demonstrate empirically in \Cref{sec:evaluation}:
\textbf{Negligible overhead (G4):} \sysname introduces minimal overhead, allowing the system to progress at the same speed as the underlying consensus protocol.
\textbf{Scalability (G5):} \sysname scales effectively with the number of auxiliary validators.
\textbf{Fault tolerance (G6):} \sysname maintains robust performance, remaining visibly unaffected by the presence of crashed auxiliary validators.

To achieve these goals, \sysname overcomes several challenges:
\textbf{(Challenge 1):} \sysname cannot implement an all-to-all communication design due to the impractical number of auxiliary validators.
\textbf{(Challenge 2):} \sysname cannot expect the classic BFT assumption to hold for these entities as they have less stake and thus less incentive to be reliable and prone to remain offline for long periods of time.
\textbf{(Challenge 3):} Auxiliary validators must participate in the consensus without causing delays, as this would undermine the key advantage of quorum-based systems. They thus cannot take actions that impact the critical path.

\section{The \sysname Design} \label{sec:design}

We present the design of \sysname and argue its properties defined in \Cref{sec:goals}.

\subsection{DAG-based consensus protocols} \label{sec:dag}

DAG-based consensus protocols operate in logical \emph{rounds}. In each round, every honest (core) validator creates a unique signed vertex. Byzantine validators may attempt to equivocate by producing conflicting vertices~\cite{dag-rider} or may abstain altogether. During each round, validators collect user transactions and vertices from other validators to construct their next vertex. Each vertex must reference a minimum number of vertices from the previous round (typically $2f + 1$~\cite{narwhal,bullshark,mysticeti}) and adds fresh transactions that do not appear in preceding vertices.

\Cref{alg:core-validator} (ignore \change{orange} for now) outlines the operations of these validators, aligning with nearly all existing structured DAG-based consensus protocols~\cite{narwhal,bullshark,shoal,shoal++,mysticeti,dag-rider,dumbo-ng,dispersedledger,sailfish,bbca-chain,fino,gradeddag,cordial-miners,wahoo,lightdag,dai2024remora} and all DAG-based systems that have been deployed in production environments~\cite{narwhal,bullshark,mysticeti,hammerhead}.

When a core validator receives a new vertex $v$, it invokes $\Call{ProcessCoreVertex}{v}$ (Line~\ref{alg:line:process-core-vertex}). The validator first downloads $v$'s causal history (Line~\ref{alg:line:sync-core-ancestors}) and verifies $v$ for validity (Line~\ref{alg:line:valid-core-vertex}), which typically involves checking signatures, validating parent vertex references, and ensuring syntactical correctness. A valid $v$ is then added to the local DAG view of the validator (Line~\ref{alg:line:add-to-dag}).

Next, the validator checks if the new vertex triggers any commits. It derives a set of \emph{leader} vertices either deterministically (in partially synchronous protocols~\cite{bullshark,shoal,mysticeti}) or by reconstructing a global perfect coin~\cite{abraham2023bingo} (in asynchronous protocols~\cite{narwhal,cordial-miners}). Using a protocol-specific decision rule, it analyzes DAG patterns to establish a total order among the leaders (Line~\ref{alg:line:commit-leaders}). If this yields a non-empty sequence, the validator linearizes the DAG into a sequence of vertices $C$ that it outputs to the application layer (Line~\ref{alg:line:linearize}). This linearization step uses a deterministic function like depth-first search over the sub-DAG defined by each leader in the sequence~\cite{dag-rider,narwhal,bullshark}.

To advance the round, the validator attempts to create a new vertex $v'$ through $\Call{TryAdvance}{\;}$ (Line~\ref{alg:line:try-advance}). This succeeds if the validator possesses enough vertices from the previous round and, in partially synchronous protocols, enough leader vertices or if a timeout has occurred. If successful, the validator adds $v'$ to its local DAG view and broadcasts it to the other validators (Line~\ref{alg:line:add-to-dag-2}). Creating a vertex may involve reliable or consistent broadcasting~\cite{dag-rider,narwhal,bullshark,shoal,shoal++}, a best-effort broadcast~\cite{mysticeti,cordial-miners}, or a hybrid of both~\cite{bbca-chain,gradeddag}.

\begin{algorithm}[t]
    \caption{Core Validator}
    \label{alg:core-validator}
    \algfontsize

    \begin{multicols}{2}
        \begin{algorithmic}[1]
            \State $T \gets \{ \; \}$ \Comment{Buffer client transactions}
            \State $\dag_c \gets \{ \; \}$ \Comment{DAG of core vertices}
            \change{\State $\dag_a \gets \{ \; \}$ \Comment{DAG of auxiliary vertices}}

            \Statex
            \Procedure{ProcessCoreVertex}{$v$} \label{alg:line:process-core-vertex}
            \State $\Call{SyncCoreAncestors}{v, \dag_c}$ \label{alg:line:sync-core-ancestors}
            \change{\State $\Call{SyncAuxAncestors}{v, \dag_a}$} \label{alg:line:sync-aux-ancestors}
            \State \require{$\Call{ValidCoreVertex}{v, \dag_c}$} \label{alg:line:valid-core-vertex}
            \State $\Call{AddToDag}{v, \dag_c}$ \label{alg:line:add-to-dag}
            \State $L \gets \Call{OrderNewLeaders}{\dag_c}$ \label{alg:line:commit-leaders}
            \If{$L \neq \perp $}
            \State $C \gets \Call{Linearize}{L, \dag_c, \change{\dag_a}}$ \label{alg:line:linearize}
            \State $\Call{OutputToApplication}{C}$
            \EndIf
            \State $\Call{TryAdvance}{\;}$
            \EndProcedure

            \Statex
            \Procedure{TryAdvance}{\;} \label{alg:line:try-advance}
            \State $v' \gets \Call{TryNewCoreVertex}{T, \dag_c, \change{\dag_a}}$ \label{alg:line:try-new-core-vertex}
            \If{$v' = \perp$} \Return \EndIf
            \State $\Call{AddToDag}{v', \dag_c}$ \label{alg:line:add-to-dag-2}
            \State $\Call{SendToCoreValidators}{v'}$ \label{alg:line:send-to-core-validators}
            \EndProcedure

            \Statex
            \change{
                \Procedure{ProcessAuxProposal}{$p$} \label{alg:line:process-aux-proposal}
                \State $\Call{SyncCoreAncestors}{v, \dag_c}$
                \State \require{$\Call{ValidAuxProposal}{v, \dag_c}$}
                \State $\sigma_p \gets \Call{Sign}{p}$
                \State $\Call{ReplyBack}{\sigma_p}$
                \EndProcedure
            }

            \Statex
            \change{
                \Procedure{ProcessAuxVertex}{$v$} \label{alg:line:process-aux-vertex}
                \State $\Call{DownloadCoreAncestors}{v, \dag_c}$
                \State \require{$\Call{ValidAuxVertex}{v, \dag_c}$}
                \State $\Call{AddToDag}{v, \dag_a}$
                \State $\Call{TryAdvance}{\;}$
                \EndProcedure
            }
        \end{algorithmic}
    \end{multicols}
\end{algorithm}
\begin{algorithm}[t]
    \caption{Auxiliary Validator}
    \label{alg:auxiliary-validator}
    \algfontsize

    \begin{algorithmic}[1]
        \State $T \gets \{ \; \}$ \Comment{Buffer client transactions}
        \State $\dag_c \gets \{ \; \}$ \Comment{DAG of core vertices}

        \Statex
        \Procedure{TryAdvance}{\;}
        \State $p \gets \Call{TryNewProposal}{T, \dag_c}$
        \If{$p = \perp$} \Return \EndIf
        \State $\{ \sigma_{(p,i)} \} \gets \Call{SendToCoreValidators}{p}$
        \State $v' \gets \Call{AssembleCertificate}{\{ \sigma_{(p,i)} \}}$
        \State $\Call{SendToCoreValidators}{v'}$
        \EndProcedure
    \end{algorithmic}
\end{algorithm}

\subsection{Vertex creation rule and commit rule} \label{sec:protocol}

\begin{figure}[t]
    \vskip -1em
    \centering
    \includegraphics[width=0.6\textwidth]{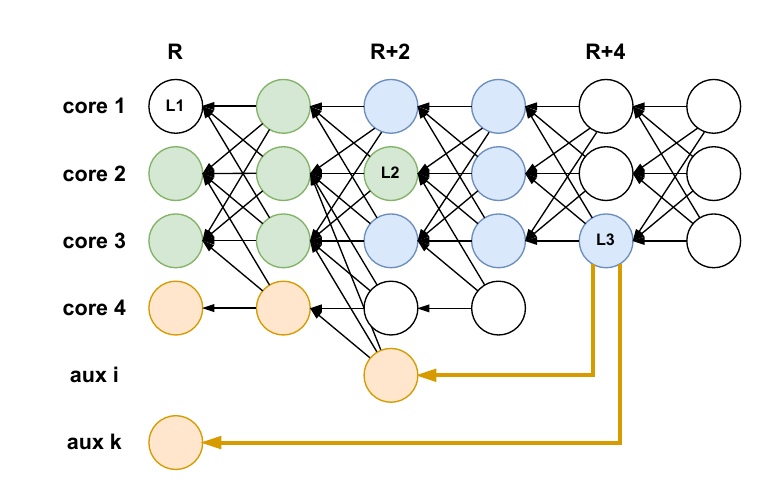}
    \caption{
        Example of \sysname execution with 4 core validators and $t_a = 2$.
    }
    \label{fig:dag}
    \vskip -1em
\end{figure}

We present the protocol for auxiliary validators and the modifications made to the core validator protocol, using \Cref{fig:dag} as an example. We denote vertices using the notation $v(author, round)$, where $author$ represents the validator that authored the vertex and $round$ indicates the round number.

\para{Auxiliary validators}
Auxiliary validators operate as full nodes, downloading the DAG generated by core validators while also collecting client transactions. \Cref{alg:auxiliary-validator} illustrates their protocol. Each auxiliary validator composes a signed \emph{proposal} containing client transactions and hash references to at least $2f + 1$ vertices created by core validators for a past round\footnote{
    Auxiliary vertices referencing core vertices too far in the past are not included in the final commit sequence, as they will be pruned by garbage collection~\cite{narwhal}.
}. They then send it to the core validators who check their validity and reply with a counter signature. The auxiliary validator then assembles a vertex, made up of the $2f + 1$ counter-signatures, and rebroadcasts it to all core validators. For example, $v_a(aux_i, R+2)$ references the vertices from round $R+1$ created by core validators $core_2$, $core_3$, and $core_4$.
This design choice solves \textbf{challenge 1} by forgoing communication between auxiliary validators and effectively leveraging the core validators as communication layer and to obtain a proof of availability for their data.

\para{Core validators}
\Cref{alg:core-validator} shows the modifications to the core validator's protocol in \change{orange}. Core validators execute $\Call{ProcessAuxProposal}{p}$ to validate, download the causal history, and counter-sign an auxiliary validator's proposal $p$. Since $p$ references $2f + 1$ core validator vertices, this verification allows core validators to synchronize any potentially missing vertices from the author of $p$. This protocol incentivizes auxiliary validators to collaborate, as inclusion of their vertices in the final commit sequence grants them a share of vertex rewards~\cite{ethereum}. The function $\Call{ProcessAuxVertices}{v_a}$ shows how core validators process a vertex $v_a$ (that is, a proposal $p$ counter-signed by $2f + 1$ core validators). Core validators add this vertex to a new map, $\dag_a$, which will later be merged into the DAG $\dag_c$ operated by core validators.
This design choice solves \textbf{challenge 2} by ensuring that malicious or unreliable auxiliary validators cannot affect the protocol once they delivered their vertex to a core validator. Since a correctly signed auxiliary vertex indicates that at least $2f + 1$ core validators possess its data and causal history, the proposed data from auxiliary validators remains highly available despite their potential unreliability.

Next, we modify the vertex creation rules: every fixed number of rounds, the core validator leader(s) (e.g., $L_3$ in \Cref{fig:dag}) must reference vertices from auxiliary validators with a joint stake totaling at least $t_a$ (see \Cref{sec:model}). In the figure, $L_3$ references $v_a(aux_i, R+2)$ and $v_a(aux_k, R)$.
This design solves \textbf{challenge 3} by allowing auxiliary validators to participate in the consensus without slowing it down. It ensures that auxiliary validators can asynchronously create vertices at a slower pace than core validators, without impacting the critical path, while still maintaining minimal required participation.

Auxiliary vertices are essentially treated as \emph{weak links}~\cite{dag-rider}, with core validators required to download them before processing a vertex. Auxiliary vertices are not used to establish the order of committed leaders but are included during the linearization step (Line~\ref{alg:line:linearize}). Designing \sysname to operate only at the linearization layer makes it compatible with nearly all DAG-based protocols: while leader ordering algorithms vary across protocols, linearization is a common procedure. The validator linearizes the vertices within the sub-DAG defined by each leader vertex through any deterministic procedure, such as a depth-first search~\cite{dag-rider}. If a vertex has already been linearized by a previous leader, the validator omits. Each leader is processed sequentially, ensuring all vertices appear in the final commit sequence in a deterministic order based on their causal dependencies.

In the example shown in \Cref{fig:dag}, the sequence of committed leaders (output from \Cref{alg:line:commit-leaders}) is ([$L_1$, $L_2$, $L_3$]).
$L_{1}$ does not define any sub-DAG (the process begins at round $R$), so only $L_1$ is added to the commit sequence.
$L_{2}$ defines a sub-DAG of the green vertices, which are linearly ordered, as e.g., $v_c(core_1,R+1)$, $v_c(core_2,R)$, $v_c(core3,R)$, $v_c(core_2,R+1)$, $v_c(core_3,R+1)$, and $L_2$.
While processing $L_3$, which defines the sub-DAG of both blue and orange vertices, the validator collects and linearizes all such vertices. As a result, although the original DAG might have excluded the core orange vertices ($v_c(core_4, R)$ and $v_c(core_4, R+1)$) and would have omitted the auxiliary vertices ($v_a(aux_i,R+2)$ and $v_a(aux_k,R)$), \sysname guarantees their inclusion in the final commit sequence. This inclusion helps core validator $core_3$ to synchronize parts of the DAG that were potential missing from its local view.

\subsection{Security analysis} \label{sec:security}
\Cref{alg:core-validator} shows that \sysname modifies the original consensus protocols only in three places: (1) Line~\ref{alg:line:sync-aux-ancestors} (to sync auxiliary vertices' ancestors), (2) Line~\ref{alg:line:linearize} (the linearization layer), and (3) Line~\ref{alg:line:try-new-core-vertex} (proposing a new vertex).

\para{Safety}
\sysname does not alter the way the underlying protocol commits leaders, ensuring safety as long as DAG linearization (Line~\ref{alg:line:linearize}) is deterministic, as is typical in existing DAG-based protocols. Safety holds regardless of the number of honest auxiliary validators.

\para{Liveness}
\sysname maintains liveness under the same network model as the base protocol by ensuring (1) core validators can synchronize missing auxiliary vertices (Line~\ref{alg:line:sync-aux-ancestors}), (2) the DAG linearization terminates (Line~\ref{alg:line:linearize}), and (3) core validators can eventually create new vertices (Line~\ref{alg:line:try-new-core-vertex}). Point (1) holds as core validators only reference certified auxiliary vertices, guaranteeing at least $f+1$ honest validators hold them~\cite{cachin2011introduction}. Point (2) is satisfied as auxiliary vertices are linearized like core ones. Point (3) follows from the assumption that at least $t_a$ auxiliary validators are honest (which we assume in \Cref{sec:model}), ensuring at least $2f+1$ core validators successfully call $\Call{ProcessAuxProposal}{\cdot}$ (Line~\ref{alg:line:process-aux-proposal}) and $\Call{ProcessAuxVertex}{\cdot}$ (Line~\ref{alg:line:process-aux-vertex}) which update their local DAGs with new auxiliary vertices, allowing the creation of new core vertices.
\section{Implementation and Evaluation} \label{sec:evaluation}
We implement\footnote{\codelink} \sysname as a fork of Mysticeti~\cite{mysticeti} and evaluate it on a geo-distributed AWS testbed. We use the same setup as the Mysticeti paper~\cite{mysticeti}\footnote{
    13 different AWS regions; \texttt{m5.8xlarge} instances (with 32 vCPUs, 128GB RAM, and 10Gbps network); 512 bytes transactions; each data point is the average latency and error bar represent one stdev; benchmarks run for multiple minutes under fixed load.
} and only test for loads up to 50k tx/s to limit costs. We set $t_a = 10\%$ (\Cref{sec:model}) and auxiliary validators propose blocks every few seconds.
We use the notation M-$X$ to indicate Mysticeti running with $X$ validators, and O-$X$-$Y$ to indicate \sysname running with $X$ core validators and $Y$ auxiliary validators.

\begin{figure}[t]
    \vskip -1em
    \centering
    \begin{minipage}{.49\textwidth}
        \centering
        \includegraphics[width=\textwidth]{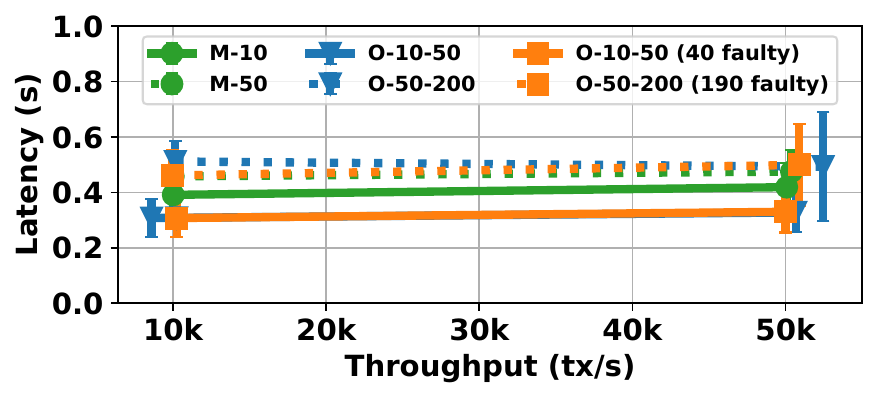}
    \end{minipage}
    \begin{minipage}{0.48\textwidth}
        \centering
        \vfill
        \includegraphics[width=\textwidth]{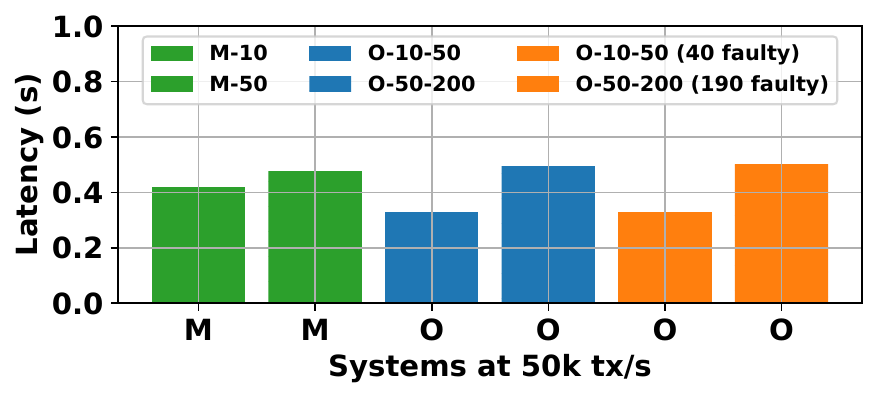}
    \end{minipage}
    \caption{
        Comparative evaluation of Mysticeti (10 and 50 validators) and \sysname (10 core + 50 auxiliary validators, 50 core + 200 auxiliary validators). Left: throughput-latency graph. Right: Latency zoom at 50k tx/s.
    }
    \label{fig:evaluation}
    \vskip -1em
\end{figure}

\Cref{fig:evaluation} shows that all system configurations maintain a latency of approximately 400ms when processing loads of either 10k tx/s or 50 tx/s. Notably, regardless of the committee size, there is no statistical difference between Mysticeti in its barebone configuration and when equipped with \sysname, confirming our claim \textbf{G4} (from \Cref{sec:goals}) that \sysname introduces negligible overhead. Additionally, \Cref{fig:evaluation} demonstrates that \sysname can scale to 200 auxiliary validators, thus validating our scalability claim \textbf{G5}. Lastly, \Cref{fig:evaluation} (orange lines) illustrates that even a large number (up to 190) of crashed auxiliary validators do not noticeably impact protocol performance, supporting our claim \textbf{G6}.
\section{Discussion and Related Work} \label{sec:related}
\sysname conceptually aligns with IOTA's open writing access~\cite{cullen2021access,muller2022tangle,iota-wiki} where all validators have writing privileges, and a ``leader'' validates blocks.
\sysname however operates fundamentally differently, as it is designed as an add-on for existing quorum-based blockchains. It draws inspiration from DagRider's weak links~\cite{dag-rider}, which include older blocks not required for leader selection (although for different reasons than \sysname), and Narwhal's vertex-creation rule~\cite{narwhal}, which ensures vertex availability before inclusion in the DAG.
Future work include the analysis of \sysname where auxiliary validators operated under the sleepy model~\cite{pass2017sleepy} to explore potential improvements in censorship resistance; whether auxiliary validators can enhance the protocol's safety for clients who trade latency, as in OFlex~\cite{malkhi2019flexible, oflex}; and whether they can aid in fork recovery when more than $f$ Byzantine core validators are present. Lastly, we leave as future work the incentive analysis and its impact on the relationship between $n_c$ and $n_a$ ($n_c < n_a$, $n_c > n_a$, or $n_c \approx n_a$).

\ifpublish
    \begin{credits}
        \subsubsection{\ackname}
        This work is supported by Mysten Labs. We thank Srivatsan Sridhar for his feedback on the paper and discussions on future work.
    \end{credits}
\fi

\bibliographystyle{splncs04}
\bibliography{references}

\end{document}